\documentclass[aps,pre,twocolumn,superscriptaddress,floatfix]{revtex4}

\usepackage{amsmath,amsfonts,amssymb}
\usepackage{graphicx}
\usepackage{algorithm}
\usepackage{algorithmic}
\usepackage{epstopdf}
\usepackage{ulem}

\let\originalleft\left
\let\originalright\right
\renewcommand{\left}{\mathopen{}\mathclose\bgroup\originalleft}
\renewcommand{\right}{\aftergroup\egroup\originalright}
\newcommand{\bra}[1]{\ensuremath{\left\langle #1\right|}}
\newcommand{\ket}[1]{\ensuremath{\left|#1\right\rangle}}

\begin{document}
\title{Hidden symmetries 
enhance quantum transport in Light Harvesting systems}

\author{Tobias Zech} 
\address{Physikalisches Institut, Albert-Ludwigs-Universitat Freiburg, Hermann-Herder-Str. 3, D-79104 Freiburg, Germany}
\author{Roberto Mulet} 
\address{Complex System Group, Department of Theoretical Physics, University of Havana, Cuba and Physikalisches Institut, Albert-Ludwigs-Universitat Freiburg,Hermann-Herder-Str. 3, D-79104 Freiburg, Germany}
\email{roberto.mulet@gmail.com}
\email{mulet@fisica.uh.cu}
\author{Thomas Wellens} 
\address{Physikalisches Institut, Albert-Ludwigs-Universitat Freiburg, Hermann-Herder-Str. 3, D-79104 Freiburg, Germany}
\author{Andreas Buchleitner}
\address{Physikalisches Institut, Albert-Ludwigs-Universitat Freiburg, Hermann-Herder-Str. 3, D-79104 Freiburg, Germany}

\date{\today}

\begin{abstract}
For more than 50 years we have known that photosynthetic systems harvest solar energy with almost unit {\it quantum efficiency}. 
However, recent experimental evidence of {\it quantum coherence} during the excitonic energy transport in photosynthetic organisms
challenges our understanding of this fundamental biological function. Currently, and despite numerous efforts, the causal connection between coherence and efficiency 
is still a matter of debate.  We show, through the study of extensive simulations of quantum coherent transport on networks, that three dimensional structures characterized by 
centro-symmetric Hamiltonians are statistically more efficient than random arrangements. 
Moreover, we demonstrate that the experimental data available for the electronic Hamiltonians of the Fenna-Mathew-Olson (FMO) complex of sulfur bacteria 
and of the crypophyte PC645 complex 
of marine algae are 
consistent with this strong correlation of centro-symmetry with quantum efficiency. 
These results show that what appears to be geometrically disordered complexes may well exhibit a hidden symmetry which enhances the energy transport between chromophores.  
We are confident that our results will motivate research to explore the properties of nearly centro-symmetric Hamiltonians in 
more realistic environments, and to unveil the role of symmetries for quantum effects in biology. 
The unravelling of such symmetries may open novel perspectives and suggest new design principles in the development of artificial devices.
\end{abstract}

\maketitle
\section{Introduction}
\label{sec:Intro}
The apparatus used by photosynthetic organisms to harvest the sun's energy is both complex and highly efficient.
Although it appears differently in different species, some features always persist: 
Photons are absorbed by pigments, usually chlorophyll or carotenoid molecules, and transmitted to a reaction 
center {\it (RC)}, where the primary chemical reactions of photosynthesis occur \cite{Voet}. 
Photosynthetic structures contain far more chlorophyll molecules than {\it RC}s, 
and the former constitute molecular networks which pass the energy of an absorbed photon to the {\it RC}, where it is trapped \cite{Voet,Hu,Fleming}.

In recent years, ultrafast optics and nonlinear spectroscopy experiments provided new insight into excitonic energy transport in 
photosynthetic organisms  \cite{Engel,Collini,Turner}. These experiments report the existence of long coherence times, 
suggesting that {\it quantum coherence } 
may play a fundamental role in the highly efficient transport of excitations, and trigger renewed theoretical interest.

A  well established strategy to simulate the quantum excitation transport in such molecular complexes is  the dynamical propagation of the electronic  
Hamiltonian in the presence of the many degrees of freedom of the environment  \cite{Ishizaki09,Mohseni08,Plenio11,Olaya08,Nalbach11,Pereverzev,Shabani11,Nalbach11a}.  
This line of thought may
suggest that proper tuning of the interaction between system and environment is responsible for the long coherence times and the efficient transport \cite{Lloyd}. 
However, 
the computationally obtained coherence times ($t_c \sim 250$~fs) \cite{Nalbach11a} are smaller than the experimentally reported ($t_c \sim 660$~fs) \cite{Engel}.
Moreover, the complexity of the quantum dynamics grows exponentially with the number of strongly coupled degrees of freedom \cite{Nalbach11a}, 
rendering approximations unavoidable, even in a full-fledged computational approach. Since, in addition, also the available experimental data characterizing the 
complexes have large uncertainties, most computational methods
use effective descriptions. This contrasts  with the fact that the non-trivial, system-specific 
information on complex quantum systems is 
inscribed into the fluctuations of characteristic observables, rather than in their average behavior \cite{gutzwiller,skourtis,scholak10,Alicki}.

On the other hand, simplified models \cite{Gilmore,Pachon, Mulken} that mimic the dynamics by those of two 
coupled two level systems interacting with the environment,  
may help to identify some minimal ingredients that must be included into a general theory. 
However,  because of their simplicity, they disregard many of the structural features of the light harvesting complexes -- 
possibly those that are most relevant for the transport efficiency in realistic systems.

Therefore, in our present contribution, we 
attempt to incorporate most of the documented properties of the underlying electronic Hamiltonians, and yet to 
outline a model as simple as possible.
We will see that 3D random networks are very likely to perform efficiently whenever their 
Hamiltonians exhibit centro-symmetry. On this basis, we 
will provide strong evidence that the transport efficiencies across light harvesting complexes indeed substantially depend on the structural properties of the 
apparently disordered macromolecular Hamiltonians. 
In particular, we will show that, specifically for the reported Hamiltonians
of the FMO8 and PC645 pigment complexes, centro-symmetry defines an evolutionary advantage.

\section{The model}
\label{sec:mod}

Inspired by the structure of the FMO network --  seven (FMO7) \cite{Fenna, Tronrud86} or eight (FMO8) \cite{Tronrud09} chromophores that are connected through  dipolar interactions -- we study the simplest possible random model which can grasp its essential ingredients: 
A small random network with $N$ sites, where the coherent transport of a single excitation is generated by the Hamiltonian
\begin{equation}
  H=\sum_{i \neq j = 1}^{N} V_{i,j}\ \sigma_+^{(j)}\sigma_-^{(i)}\, .
  \label{eq:ham}
\end{equation}
Here, $\sigma_+^{(j)}$ and $\sigma_-^{(i)}$ mediate excitations and de-excitations of sites $j$ and $i$, from the local electronic ground state to the local excited state, and vice versa.
The excitation transfer $\sigma_+^{(j)}\sigma_-^{(i)}$ from site $i$ to site $j$ has a strength  $V_{i,j}=V_{j,i}=\alpha/r_{i,j}^{3}$, consistent with an isotropic 
dipolar interaction, with $r_{i,j} = |\vec r_i-\vec r_j|$, and the $\vec r_j$ the position vectors of individual sites. Input and output sites define the pole of a sphere of diameter $d$. 
The positions of the remaining molecular sites are  randomly chosen within this sphere,  what induces a random distribution of the remaining $V_{i,j}$. Networks for which one of the distances $r_{i,j}$  
was smaller than $d/100$ were discarded in the  analysis, in order to avoid singular coupling strengths \cite{diss_scholak}.
The excitation initially is injected at the input site $|\text{in}\rangle = |1\rangle$, from where it is to be transferred to the output site $ | \text{out}\rangle = |N\rangle$.
The coupling constant $V_{1,N} = \alpha/d^3$ between these two sites sets the natural time-scale of the dynamics induced by $H$.
The intuition is that the additional sites, 
if properly placed, can mediate a multitude of transition amplitudes from input to output, which interfere {\it constructively} upon transmission, and thus ease the excitation transfer.
A network will be considered {\it efficient} if the initial excitation is  transferred to the output site in a time significantly smaller than the Rabi coupling time $T = \pi/2|V_{1,N}|$ 
between $\ket{\rm in}$ and $\ket{\rm out}$, with high probability. 
For a quantitative assessment, we define the figure of merit \cite{Scholak,Scholak11}
\begin{equation}
  {\mathcal P}= \max_{t \in [0, \mathcal T]} |\langle N | e^{-i H t}  | 1 \rangle|^2\, ,\,
  \text{with} \;  \mathcal{T} = 0.1 \times \pi/2|V_{1,2}| \, .
  \label{eq:transportefficiency}
\end{equation}
Indeed, we showed earlier that ${\mathcal P}$ fluctuates strongly with different realizations of the random molecular network, and reaches large values close to unity with finite probability.
However, while it is therefore natural to ask for necessary and/or sufficient conditions on the networks's structure such as to guarantee large 
values of ${\mathcal P}$, no {\it design principles} were so far identified. 
This is the purpose of our present contribution. 

\section{Results}

While near-optimal random networks -- in the sense of giving rise to large values of ${\mathcal P} \simeq 1$ -- do not exhibit apparent symmetries in their 3D geometry, it was 
observed that the individual sites' populations indeed do display a near-symmetric structure on the time axis, under exchange of input and output site, as well as of properly defined 
pairs of intermediate sites (see Fig.~2 in \cite{Scholak11}).
This must be inherited from an exchange symmetry of pairs of two-site coupling matrix elements of the underlying Hamiltonian, and evokes an 
analogy with $N\times N$ {\it centro-symmetric} matrices, which are defined by $H_{i,j} = H_{N-i+1,N-j+1}$, i.e., $JH = HJ$, where $J$ is the exchange matrix, $J_{i,j}= \delta_{i,N-j+1}$, that 
exchanges site 1 with site $N$, 2 with $N-1$, etc. 
This type of Hamiltonians 
is known 
to be tunable towards optimal 
excitation transfer \cite{Cantoni,Albanese,Kay}.  
\begin{figure}[htbp] 
   \centering
  	\includegraphics[width=8cm]{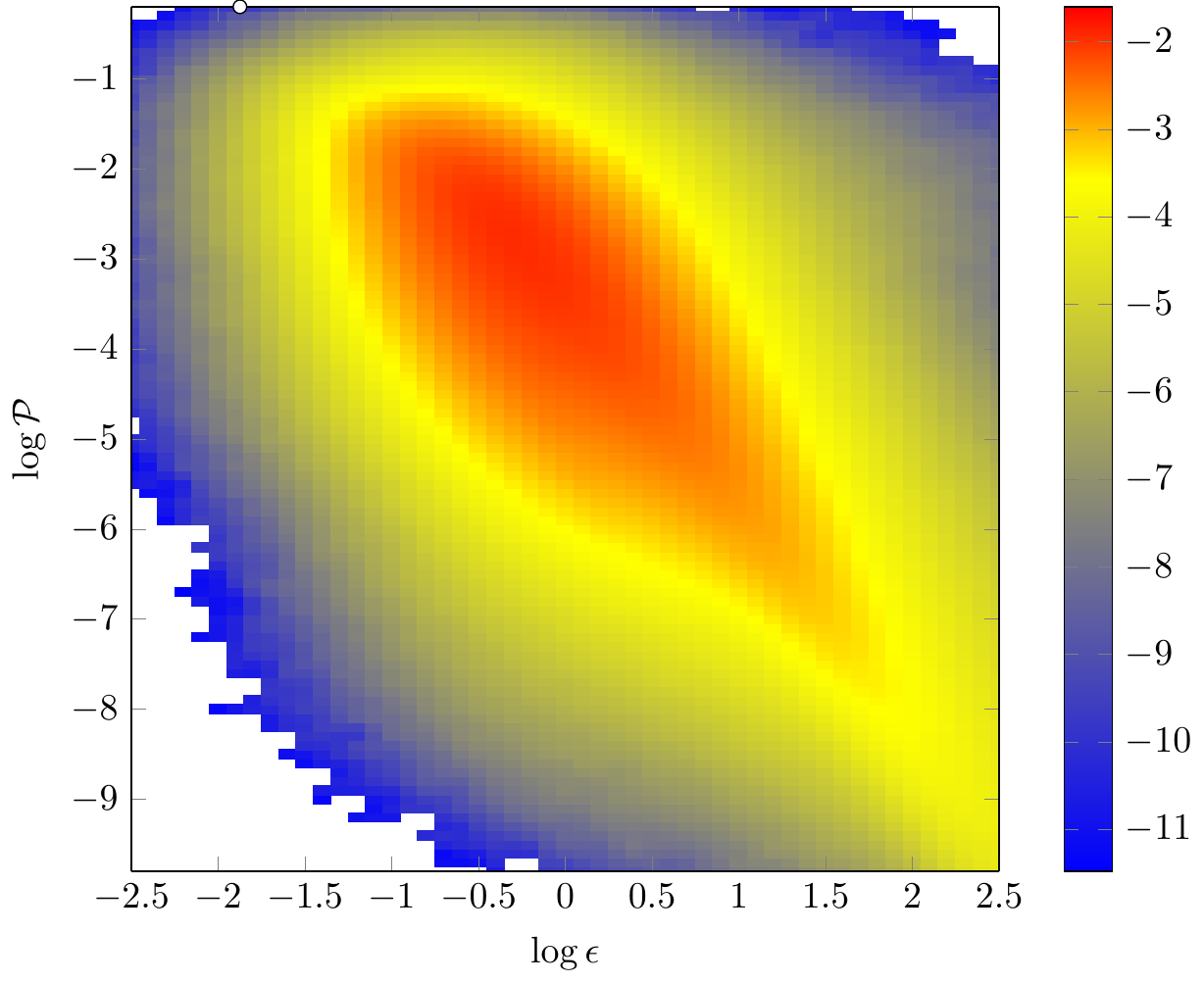}
  \caption{
  Correlation between efficiency $\mathcal{P}$ and centro-symmetry $\epsilon$ of $10^8$ randomly generated 3D networks composed 
  of seven sites.
  The color code  indicates the logarithm of the probability density of a pair  $(\log (\epsilon),\log(\mathcal{P}))$. The white circle close to the top left corner represents 
  a quasi optimal 
  configuration found by a genetic algorithm.}
   \label{fig:effvssym}
\end{figure}
We now quantify the transfer efficiency (\ref{eq:transportefficiency}) of a given Hamiltonian, and correlate this with 
the Hamiltonian's centro-symmetry, measured  by $\epsilon =\min_S ||H - J^{-1}HJ||$, where $|| \dots ||$ denotes the 
Hilbert-Schmidt norm, i.e.,  $||A||=\sqrt{{\rm Tr}A^\dagger A}$ \cite{Reed}. The quantity $\epsilon$ measures the 
root-mean square deviation of a network Hamiltonian from its image under $J$, minimized over all possible  permutations $S$ of the intermediate sites $2,\ldots, N-1$.
Figure \ref{fig:effvssym} shows the correlation plot on a triple logarithmic scale, for $N=7$, and sampling $10^8$ independent random realizations of the network.

A very strong and, given the variance of ${\mathcal P}$ and $\epsilon$ over several orders of magnitude, unambiguous correlation between centro-symmetry 
(small values of $\epsilon$) and transfer efficiencies is evident. 
So is the rather low probability to generate a high efficiency molecular conformation by random sampling. 
On the other hand, the observed strong statistical correlation implies that an iterative search for highly efficient networks in conformation space will with very high 
probability move towards higher degrees of centro-symmetry, and hence acts as an evolutionary {\it funnel}.
This is confirmed by a genetic algorithm \cite{GA} 
 that optimizes the structure of the networks to achive maximum efficiency: Starting from random networks, located with high probability close to the global maximum of the 
 distribution in Figure \ref{fig:effvssym}, i.e.\ approx. at $\log(\epsilon) \simeq -0.5$, $\log(\mathcal{P}) \simeq -3$, the algorithm generates 
 configurations with almost unit efficiency and very low centrosymmetry.  An exemplary output
 configuration is indicated by the white circle in Figure \ref{fig:effvssym}:
While the deviation from the centro-symmetry is reduced by approximately $1.5$ orders of magnitude, the network's efficiency increases by about $3$ orders of magnitude.

\begin{figure}[htbp] 
   \centering
   \includegraphics[width=8cm]{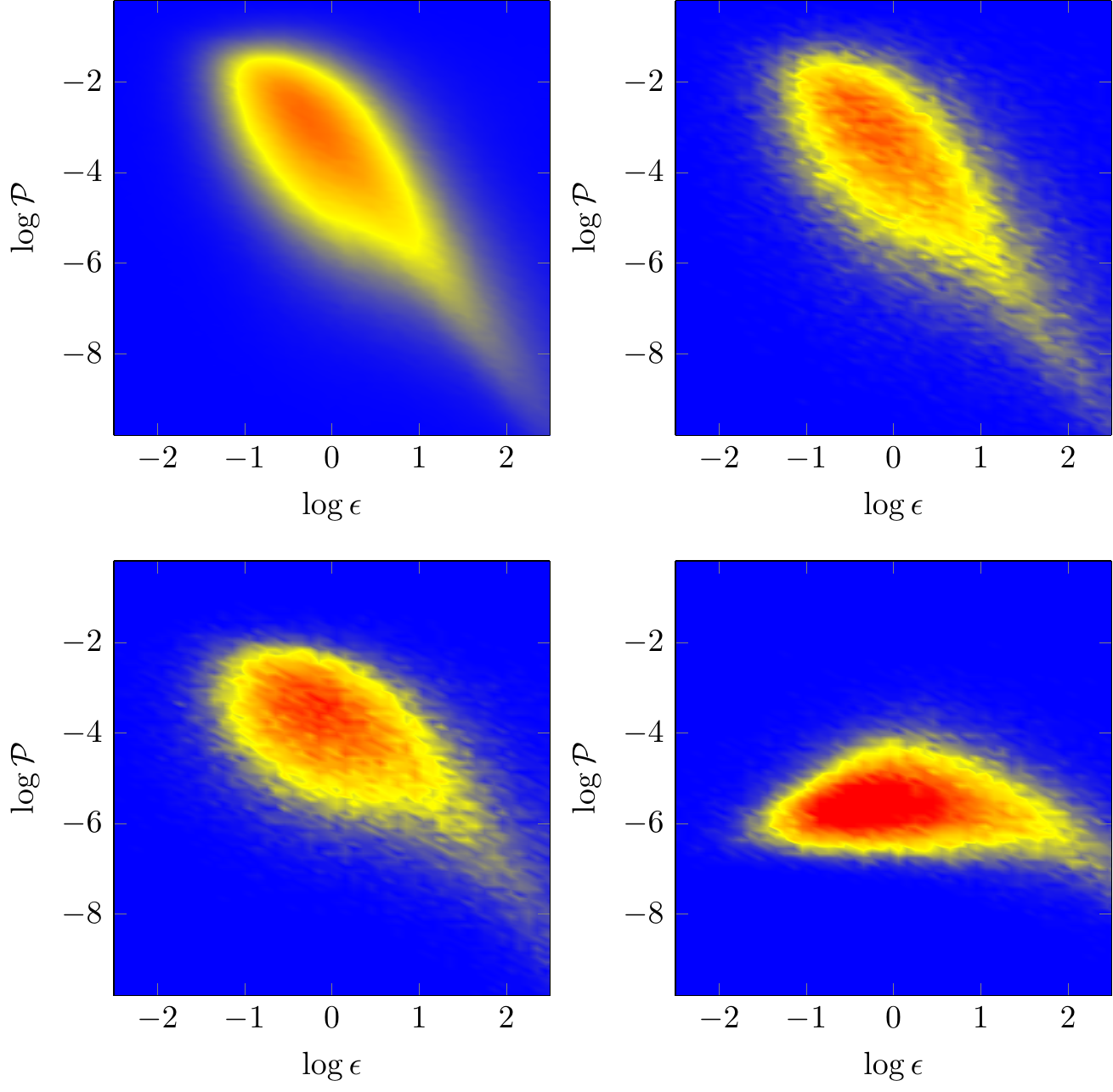}
   \caption{Histogram for the open system dynamics of mirrorsymmetry $\epsilon$ vs.
transport efficiency $\mathcal{P}$, for different values of the dephasing rate in the site basis, 
$\gamma = $ (upper left) $0$, (right) $1$, (lower left) $10$, (right) $100$, in units of the characteristic time scale $T=\pi/2|V_{\textrm{in},\textrm{out}}|$. The color code  indicates the probability density of a pair  $(\log (\epsilon),\log(\mathcal{P}))$. The histograms are generated from $10^5$ random networks for $\gamma>0$, and $10^8$ for $\gamma=0$ (as in Figure~\ref{fig:effvssym}, but on linear scale). The correlation between transfer efficiency and centrosymmetry is less pronounced for larger values of $\gamma$.}
   \label{fig:effvssymdeph}
\end{figure}

That the observed correlation between ${\mathcal P}$ and $\epsilon$ is in itself of quantum origin is suggested by our initially formulated 
intuition on the role of the intermediate sites which mediate the excitation transfer from $\ket{\rm in}$ to $\ket{\rm out}$.
This is underpinned by inspection of the analogously generated correlation plots in Figure \ref{fig:effvssymdeph}, though in the presence of 
environment-induced dephasing locally at each individual molecular site.
The excitation transfer dynamics is then no longer described by the unitary evolution alone, but rather by a 
Lindblad equation $\dot{\varrho}(t) = - \mathrm{i} \, [H, \varrho(t)]  + L_{\mathrm{deph}}(\varrho(t))$, with the Lindblad term
\begin{align}
  L_{\mathrm{deph}}(\rho) = - 4 \gamma \sum_{i \neq j = 1}^7
  \ket{i} \bra{i}  \varrho \ket{j} \bra{j}\, ,
  \label{eq:linddeph}
\end{align}
and $\ket{\rm i}$ and $\ket{\rm j}$ the N-site electronic state where the excitation is located at the $\text{i}^{th}$, respectively $\text{j}^{th}$ site. 
As the dephasing rate $\gamma$ is tuned from zero over one and ten to hundred incoherent events per Rabi coupling time $T$, the 
distribution in the correlation plot essentially rotates from a strongly increasing efficiency with increasing centro-symmetry to an 
essentially centro-symmetry-independent distribution, at the largest dephasing rates.
However, note that the correlation prevails even for relatively large dephasing rates of $\gamma = 10 T^{-1}$, what corresponds to 
one incoherent event within the time interval that we chose to define ${\mathcal P}$ in (\ref{eq:transportefficiency}).

We now need to validate the relevance of our above findings for the actual design of light harvesting complexes as they abound in nature. Clearly,
the strong correlation between transfer efficiency and centro-symmetry of the network as evident from Figs. 1 and 2 
was deduced from unrestricted, uniform statistical sampling over a random Hamiltonian of the
form (\ref{eq:ham}). However, actual biological functional units are sufficiently well characterized (by spectroscopic means) to locate them in a comparably 
small sub-volume of conformation space. 
Therefore, we now focus on the trade-off between centro-symmetry and transfer efficiency using
published 
data of the FMO8 protein \cite{Schmidt11} and the PC645 \cite{Huo11} light harvesting complexes (of sulfur bacteria and marine algae, respectively) as our point of departure. Both these complexes 
are represented by a network of eight dipole-coupled two-level systems, and thus comply with our abstract model from above.

In order to define an unbiased benchmark in comparison to which we can quantitatively assess the centrosymmetry properties of the actual  FMO8 and the PC645 Hamiltonian, we use
an ensemble of random Hamiltonian matrices constructed as follows:
the off-diagonal elements of these random Hamiltonians are modeled as identically and independently distributed random variables, which are all chosen from the same 
Gaussian distribution. The mean value and standard deviation of this distribution is  extracted as the mean and as the standard deviation of the
non-diagonal elements of the published data for the FMO8 and PC645 complex, respectively (see the complementary material in Sec.~\ref{sec:hamiltonians}).
 In order to avoid centro-symmetry to be masked by the biological complexes' on-site energies induced by their coupling to (environmental) background degrees of freedom, we in addition set all
diagonal matrix elements equal to their average value.

The distribution of the centro-symmetry
$\epsilon$ with respect to the input  and output sites thus obtained is represented by the green histograms, for FMO8 and PC645, respectively, in Figure 3, and 
defines our unbiased benchmark distribution. For the FMO8, the input site is the site recently added to the structure of the protein, and the output site is usually named ``site 3" in the 
biological literature \cite{Schmidt11}. For the PC645, we identify the DBVc and PCBd82 molecules with the input and output sites, respectively \cite{Huo11}.

\begin{figure}[htbp] 
   \centering
      \includegraphics[width=4cm]{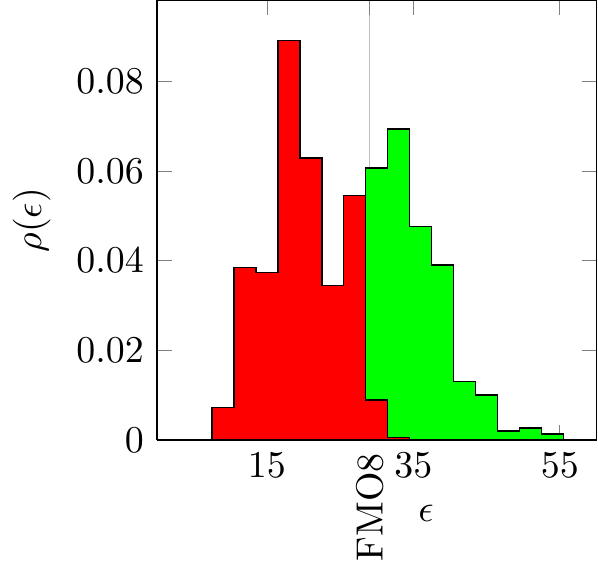}
      \includegraphics[width=4cm]{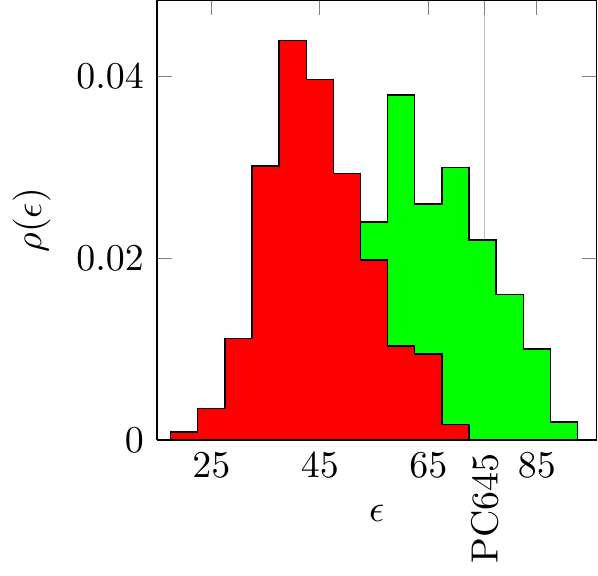}
   \caption{Green: Centro-symmetry distributions for Gaussian random ensembles 
   with mean and standard deviation extracted from the original FMO8 (left) and PC645 (right) 
      structure data \cite{Schmidt11,Huo11}. 
   Red: Centro-symmetry distributions for efficient systems obtained by evolutionary optimization of the FMO8 and PC645 Hamiltonians. The vertical lines indicate the 
   centro-symmetry of the original FMO8 and PC645 Hamiltonians, respectively.}
   \label{fig:histFMO}
\end{figure} 
The vertical lines in Fig.~3 indicate the centro-symmetries of the effective FMO8 and PC645 Hamiltonians, respectively,
extracted from the published data.
 Since, in both cases, the complexes'
centro-symmetries are located in the bulk of the unbiased distribution, they do not exhibit any significant symmetry properties. This is consistent with 
the time evolution of the single site populations as presented in Figure 4, which clearly show very unsatisfactory coherent transfer efficiency, for both molecular
networks. 
Indeed, as we have checked, {\it all} elements of the random ensemble display similarly mediocre efficiencies.
\begin{figure}[htbp] 
   \centering
   \includegraphics[width=8cm]{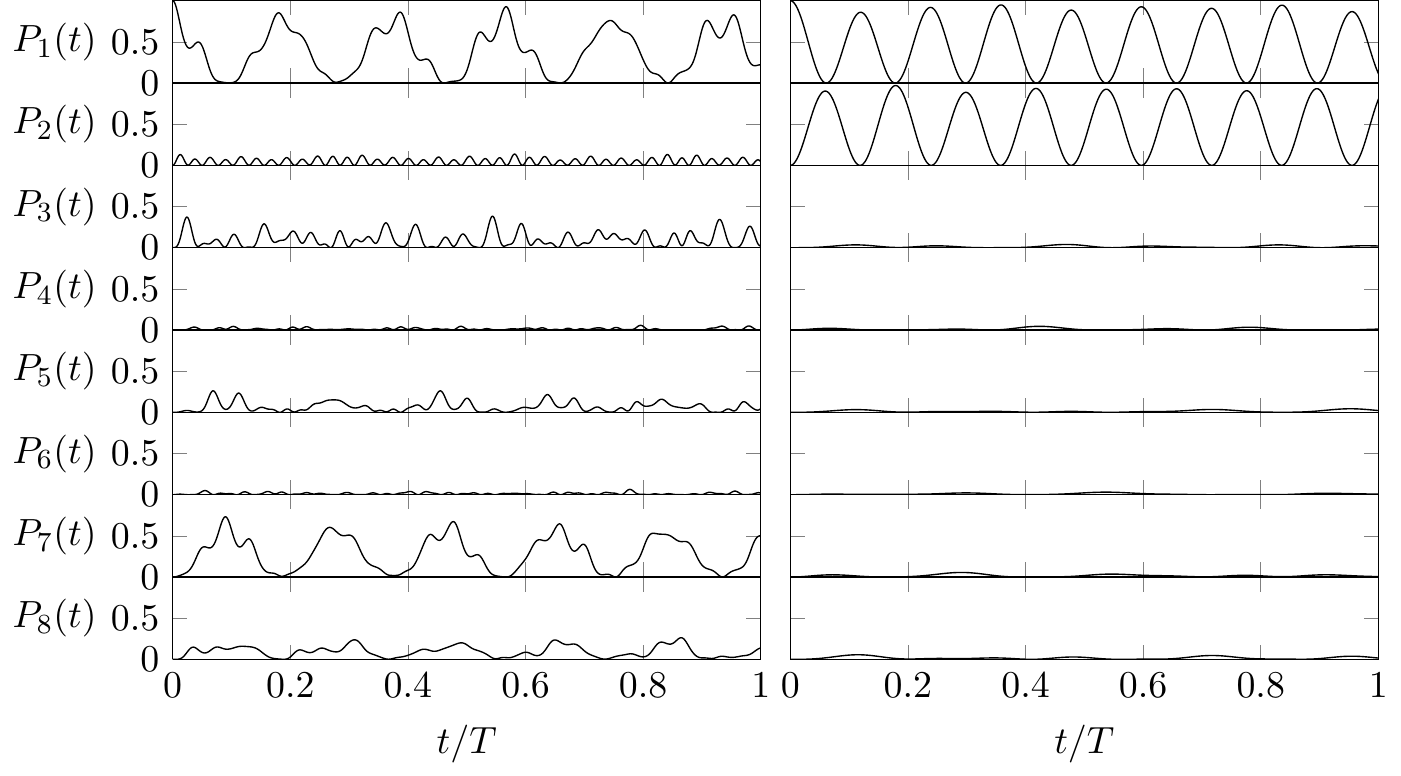}  
   \caption{Single site populations 
      of the FMO8 (left) and PC645 (right) light harvesting complexes, with inter-site coupling matrix elements extracted from published structure data \cite{Schmidt11,Huo11}, and 
   on-site energies replaced by their mean. Initially, the excitation is located at site 1. Clearly, the transfer efficiency towards site 8 is mediocre, for FMO8 as well as for PC645.
   Site labels 1 and 8 correspond to sites 1 and 3 for FMO8 \cite{Schmidt11}, and to DBVc and PCBd81 for PC645 \cite{Huo11}, respectively.}
   \label{fig:dynfmo}
\end{figure} 

We now assess how close FMO8 and PC645 structures are, in conformation space, to more centro-symmetric conformations, with higher transfer efficiencies.
For this purpose, we generate a new ensemble of network Hamiltonians, by repeated execution of N=10000 iterations of a 
simple genetic algorithm (see Sec.~\ref{sec:genetic}), seeded by the 
published network Hamilonians \cite{Schmidt11,Huo11} as indicated by the vertical lines in Figure 3. This 
results in the red histograms in Figure 3, which are clearly shifted towards higher centro-symmetries (i.e., smaller values of $\epsilon$), 
by more than the original benchmark distributions' widths. 
We verified that all the thus optimized Hamiltonians mediate essentially perfect excitation transfer. 
Moreover, even the {\it average} Hamiltonian obtained from an equally weighted sum of the optimized ensemble  
does so! This is clearly demonstrated by Figure 5, where the time dependence of the individual site populations is monitored, as generated 
by this average Hamiltonian.
\begin{figure}[htbp] 
   \centering
	\includegraphics[width=8cm]{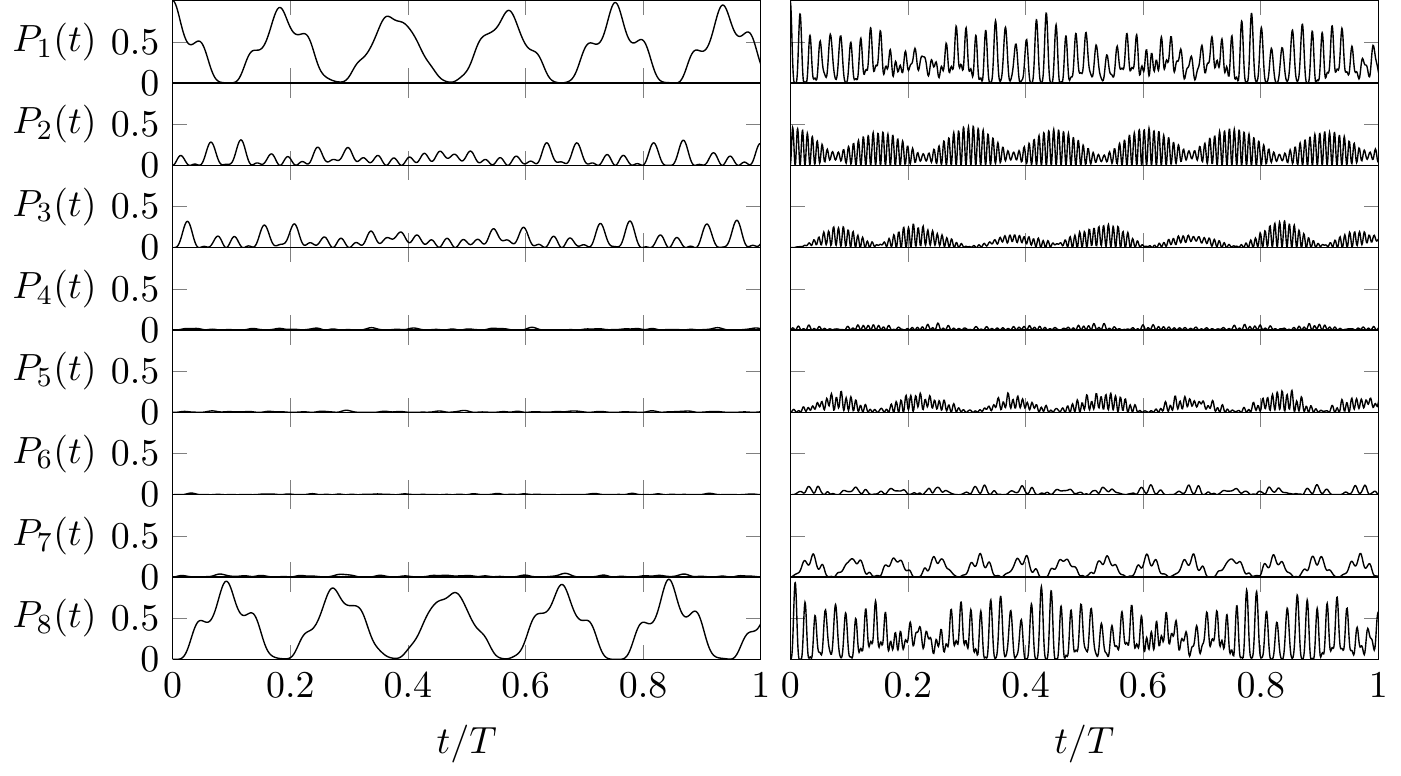}
   \caption{Single-site populations
   for genetically optimized 
   FMO8 (left) and PC645 (right) light harvesting complexes, with inter-site coupling matrix elements given by an equally weighted average 
   over $10^3$ optimized Hamiltonians generated 
   by the genetic algorithm. Site labeling as in Fig.~\ref{fig:dynfmo}. Essentially optimal transfer efficiencies are achieved in both cases.
    }
   \label{fig:dynfmo8ga}
\end{figure} 
Since different, individually optimized Hamiltonians generally do not commute, this is a highly nontrivial result which underpins, in 
particular, that evolutionary optimization seeded by the FMO8 and PC645 structures 
converges into essentially optimal transfer efficiencies, in a statistically robust sense. 
Furthermore, this observation highlights 
close connection between centro-symmetry and efficiency, since a sum of centro-symmetric Hamiltonians remains centro-symmetric. Finally, inspection of the thus obtained average optimal
Hamiltonian's matrix elements shows that most of them 
can be obtained by weak perturbation of those as inferred spectroscopically (see Sec.~\ref{sec:hamiltonians}).

In contrast to the above results when starting out from the FMO8 and PC645 structures, 
the very same genetic optimization
leaves the distributions as indicated by the green histograms in Figure 3 essentially unaffected, with correspondingly bad transfer efficiencies, when seeded by any of the elements
of our random benchmark ensemble.
This markedly distinct behavior of the genetic optimization for FMO8 and PC645 on the one hand, and of random benchmark Hamiltonians on the other therefore strongly 
suggests 
centro-symmetry as a hidden design principle.

\section{Conclusions}

Symmetry is a key concept in physics,  relevant in fields as diverse as High Energy Physics and Condense Matter. In our present work, we provide strong evidence to support the 
hypothesis
that  hidden symmetries in the Hamiltonian of apparently disordered biological structures foster quantum capabilities. In particular, we have shown that in random artificial 
networks, but also in the biologically relevant light harvesting complexes FMO8 and 
PC645, (nearly) centro-symmetric Hamiltonians define a robust design principle to mediate coherence-induced, efficient quantum transport. Note that this hidden symmetry property
could only be extracted by a statistical analysis of suitably constructed matrix ensembles, gauged against experimentally inferred average data. This is not surprising, since complex quantum systems typically exhibit very intricate spectral properties, with in general only small embedded regions (e.g. on the energy axis 
or with respect to some control parameter(s)) of   
locally well-defined quantum numbers.
In contrast to the global symmetries of integrable quantum systems, the associated (local) symmetries of complex systems are typically not given by explicitly defined 
integrals of motion, and this is why they are hard to detect -- and yet may manifest a dramatic and robust impact on the systems' dynamical properties.
We trust that our findings may define 
a good starting point for the design of new devices for efficient energy transduction \cite{Sperling}, and are confident that the here unveiled 
conspiracy of symmetry and quantum 
coherence in an apparently disordered system may be 
instrumental in other biological systems \cite{Bothma}.

\section{Acknowledgments}
We are indebted to Irene Burghardt, Rienk van Grondelle, Shaul Mukamel, Torsten Scholak, Greg Scholes and Mattia Walschaers for stimulating discussions and useful 
insights. R.M. acknowldeges support by the Alexander von Humboldt Foundation through a research fellowship.

\onecolumngrid
\section{Supplementary material}

\subsection{Genetic Algorithm}
\label{sec:genetic}
To iteratively optimize the transfer efficiency we proceed 
as follows:
\begin{enumerate}
\item[Step 1:] Read the original Hamiltonian: $H^{*}$.
\item[Step 2:] Starting from $H^{*}$, generate $M$ configurations by randomly perturbing all matrix elements $H_{ij}$ of $H^*$, according to a Gaussian distribution with 
standard deviation 
$\sigma$, initially set to $\sigma_0=H_{ij}/10$.
\item[Step 3:] From these configurations (including $H^*$), choose the one with the largest efficiency. This configuration defines the new $H^{*}$.
\item[Step 4:] Repeat steps 2-4 until the maximum number $N$ of iterations is reached. At the $n$th iteration, perturb the matrix element $H_{ij}$ by a Gaussian distribution
with standard deviation $\sigma_n=\sigma_0/n$.
\end{enumerate}

In our optimization of the FMO8 and PC645 structures, $M=100$ and $N=10000$ always achieve convergence. 
Note that the 
algorithm only optimizes with respect to the transfer efficiency, and not with respect to the centrosymmetry. 

\subsection{Hamiltonians}
\label{sec:hamiltonians}

\begin{figure}[htb]
{
\begin{align*}  
\left(
\begin{array}{cccccccc}
\ldots & -12.7/-11.1 & 35.9/39.5 & 3.6/7.9 & -1.6/-1.6 & 4.8/4.4 & -7.0/-9.1 & 1.4/1.4\\
\ldots& \ldots& -12.8/-12.2& 6.4/5.7& -96.4/-62.3& -4.9/-4.6& 21.9/35.1& 3.8/3.4 \\
\ldots& \ldots & \ldots & -96.2/-98.0 & -6.0/-5.9 & 7.9/7.1 & -12.4/-15.1 & 2.8/5.5\\
\ldots& \ldots & \ldots & \ldots & 7.9/7.6&  1.5/1.6& 12.7/13.1 & 35.8/29.8\\
\ldots& \ldots & \ldots& \ldots& \ldots & -40.9/-64.0 & -16.2/-17.4 & -9.3/-58.9\\
\ldots& \ldots & \ldots& \ldots& \ldots& \ldots & 134.9/89.5 & -1.2/-1.2\\
\ldots& \ldots & \ldots& \ldots& \ldots& \ldots& \ldots & -7.9/-9.3\\
\ldots& \ldots & \ldots & \ldots & \ldots & \ldots & \ldots &  \ldots
\end{array}
\right)
\end{align*}
}
\caption{
Off-diagonal elements of the average Hamiltonian after optimizing the FMO8 structure with a genetic algorithm (left), and of the original, spectroscopically inferred 
FMO8 Hamiltonian (right) \cite{Schmidt11}.\label{fig:hamiltonian_fmo}}
\end{figure}
\begin{figure}[htb]
{
\begin{align*} 
\left(
\begin{array}{cccccccc}
\ldots & -10.6/-46.8 & 106.4/319.4 & -5.7/-9.6& -23.8/-43.9 & 18.4/20.3 & 3.2/25.3 & -0.7/-20.0\\
\ldots& \ldots& 14.4/21.5& -17.1/-15.8& 56.1/53.8& 10.9/11.0& 29.2/29.0& 5.1/48.0 \\
\ldots& \ldots & \ldots & 30.2/43.9 & 7.1/7.7 & 16.3/30.5 & 23.3/29.0 & 105.4/48.0\\
\ldots& \ldots & \ldots& \ldots & 4.5/4.3&  -154.4/-86.7& -3.0/-2.9 & 14.5/49.3\\
\ldots& \ldots & \ldots& \ldots& \ldots & 3.5/3.4 & 108.4/86.2 & -20.2/-14.7\\
\ldots& \ldots & \ldots& \ldots& \ldots& \ldots & 7.9/7.8 & 11.4/10.0\\
\ldots& \ldots & \ldots& \ldots& \ldots& \ldots& \ldots & -7.6/-10.7\\
\ldots& \ldots & \ldots& \ldots & \ldots & \ldots & \ldots &  \ldots
\end{array}
\right)
\end{align*}
}
\caption{
Off-diagonal elements of the average Hamiltonian after optimizing the PC645 structure with a genetic algorithm (left), and of the original, spectroscopically inferred 
PC645 Hamiltonian (right) \cite{Huo11}.\label{fig:hamiltonian:pc645}}
\end{figure}
A comparison of the off-diagonal elements of the average optimized Hamiltonian with the original FMO8 and PC645 Hamiltonians as given in the literature is 
shown in Figs.~\ref{fig:hamiltonian_fmo} and \ref{fig:hamiltonian:pc645}. 
Using the standard notation from the literature, the rows in the Hamiltonian of Fig.~\ref{fig:hamiltonian_fmo} (FMO8) correspond to the sites with labels 8, 1, 2, 4, 5, 6, 7, and 3.
Likewise, the rows in the Hamiltonian of Fig.~\ref{fig:hamiltonian:pc645} (PC645) correspond to sites labeled by DBVc, PCBc82, DBVd, MBVa, MBVb, PCBc158, PCBd158, and PCBd82. 
Mean values $\overline{V}$ and standard deviations $\Delta V$ which define the random ensembles represented by the green histograms 
in Fig.~\ref{fig:histFMO} are derived from the data in Figs.~\ref{fig:hamiltonian_fmo}, \ref{fig:hamiltonian:pc645} as:
$\overline{V} =  2.6$, $\Delta V = 34.5$ (FMO8), and $\overline{V} = 21.3$, $\Delta V =  67.0$ (PC645), respectively.

\end{document}